\documentclass[a4paper]{jpconf}

\usepackage{a4wide}
\usepackage{amssymb}
\usepackage{amsmath}
\usepackage{amsthm}
\usepackage{latexsym}
\usepackage{epsfig}
\usepackage{graphicx}
\usepackage{axodraw}

\def\de{\partial}
\def\a{\alpha}
\def\b{\beta}
\def\g{\gamma}

\def\d{\delta}
\def\D{\Delta}

\def\la{\lambda}
\def\La{\Lambda}
\def\k{\kappa}
\def\m{\mu}
\def\n{\nu}
\def\r{\rho}

\def\s{\sigma}

\def\t{\tau}

{\rm }

\def\vf{\varphi}

\def\de{\partial}

\newcommand{\be}{\begin{equation}}
\newcommand{\ee}{\end{equation}}
\newcommand{\bea}{\begin{eqnarray}}
\newcommand{\eea}{\end{eqnarray}}
\newcommand{\beqar}{\begin{eqnarray*}}
\newcommand{\eeqar}{\end{eqnarray*}}
\newcommand{\eg}{{\it e.g.,}\ }
\newcommand{\ie}{{\it i.e.,}\ }
\newcommand{\bdm}{\begin{displaymath}}
\newcommand{\edm}{\end{displaymath}}

\newcommand{\reef}[1]{(\ref{#1})}

\begin{document}

\title{Conical singularities regularized in warped
six-dimensional flux compactification and their induced brane
cosmology}

\author{Vassilios Zamarias}

\address{Department of Physics, National Technical University of
Athens, \\
Zografou Campus GR 157 73, Athens, Greece\\ and\\
Department of Physical Sciences, Hellenic Military Academy, \\
Vari, 16673, Attica, Greece}

\ead{zamarias@central.ntua.gr}

\begin{abstract}

We discuss the regularization of codimension-2 singularities in
warped six-dimensional Einstein-Maxwell axisymmetric models by
replacing them by codimension-1 branes of a ring form, situated
around the axis of symmetry. Further we consider the case of one
capped regularized conical brane of codimension one and study the
cosmological evolution which is induced on it as it moves in
between the known {\it static} bulk and cap solutions. We present
the resulting brane Friedmann equation which gives a dominant
five-dimensional $\rho^2$ energy density term at high energies and
a term linear to the energy density at low energies with, however,
negative coefficient in the small four-brane radius limit (\ie
with negative effective Newton's constant).

\end{abstract}


\section{Introduction}

Cosmology in theories with branes embedded in  extra dimensions
has been the subject of intense investigation during the last
years. The most detailed analysis has been done for braneworld
models in five-dimensional space \cite{5d1}. The effect of the
extra dimension can modify the cosmological evolution, depending
on the model, both at early  and late times. The cosmology of this
and other related models with one transverse to the brane extra
dimension (codimension-1 brane models) is well understood
\cite{RScosmology,5dcosmo,induced,indcosmology,GB1,GBcorrection,idg+gb}
(for a review see \cite{reviews}).

Braneworld models can also be extended in higher than
five-dimensions. We can consider a $(n+1)$ brane embedded in
$(n+2)$ spacetime (codimension-1 models) or a $(3+1)$ brane
embedded in $(n+1)$ spacetime (codimension-$n-3$ models). Six or
higher-dimensional braneworld models of codimension-1 are
considered as generalizations of the Randall-Sundrum model
\cite{HighCodim1}.

Six-dimensional brane world models are rather interesting to be
studied for a number of reasons. Firstly the large
(sub-millimeter) extra dimensions proposal for the resolution of
the electroweak hierarchy problem was provided \cite{AADD} in the
framework of brane theories with two extra transverse dimensions.
Therefore the study of extra dimensional theories, and string
theory in particular, become relevant to low energy phenomenology
(colliders as well as astrophysical and cosmological observations)
and is testable in the very near future. The different
possibilities of realizing a brane world model in six dimensions
must therefore be studied to allow for a comparison with
experiment.

Another attractive motivation has been the proposal to ameliorate
the cosmological constant problem, using codimension-2 branes (for
a recent review on the subject see \cite{koz}). These branes have
the interesting property that their vacuum energy instead of
curving their world-volume, just introduces a deficit angle in the
local geometry \cite{CLP}. Models with this property which exhibit
no fine-tuning between the brane and bulk quantities have been
known as self-tuning (for early attempts to find similar models in
five dimensions see \cite{5d}). Such self-tuning models with flux
compactification \cite{6dflux,susy} have been extensively looked,
but the flux quantization condition always introduces a fine
tuning \cite{fluxquant}, unless one allows for singularities more
severe than conical \cite{noncon}. Alternative sigma-model
compactifications have been shown to satisfy the self-tuning
requirements \cite{6dsigma}. However, the successful resolution of
the cosmological constant problem would also require that there
are  no fine-tuning between bulk parameters themselves. No such
self-tuning model has been found yet with all these properties.

A further motivation in studying such models with codimension-2
branes is that gravity on them is purely understood. The
introduction of matter (\ie anything different from vacuum energy)
on them, immediately introduces malicious non-conical
singularities \cite{Cline}. A way out of this problem is to
complicate the gravity dynamics by adding a Gauss-Bonnet term in
the bulk or an induced curvature term on the brane, in which case
the singularity structure of the theory is altered and non-trivial
matter is allowed \cite{GB}. However, the components of the
energy-momentum tensor of the brane and the bulk are tuned
artificially and the brane matter is rather restricted
\cite{GBcon}. Alternatively, one can regularize the codimension-2
branes by introducing some thickness and then consider matter on
them \cite{regular}. For example,  one can  mimic the brane by a
six-dimensional vortex (as \eg in \cite{uzpe}), a procedure which
becomes a rather difficult task if matter is added on it.

Even thought we do not fully understand black hole solutions in
codimension-2 braneworlds they have been extensively discussed in
\cite{Charmousis:2008bt,Kaloper:2006ek,Kiley:2007wb,CPTZ}. Lately
a six-dimensional black hole localized on a 3-brane of
codimension-2~\cite{Kaloper:2006ek,Kiley:2007wb} was proposed.
However, it is not clear how to realize these solutions in the
thin brane limit where high curvature terms are needed to
accommodate matter on the brane. The localization of a black hole
on the brane and its extension to the bulk is a difficult task. In
codimension-1 braneworlds the first attempt was to consider the
Schwarzschild metric and study its black string extension into the
bulk~\cite{Chamblin:1999by}. which is unfortunately unstable to
classical linear perturbations~\cite{BSINS}. Since then, several
authors have attempted to find the full metric using numerical
techniques \cite{BHNUM}. Analytically, the brane metric equations
of motion were considered with the only bulk input coming from the
projection of the Weyl tensor~\cite{SMS} onto the brane. Since
this system is not closed because it contains an unknown bulk
dependent term, assumptions have to be made either in the form of
the metric or on the Weyl term~\cite{BBH}. In codimension-2
braneworlds, recently black holes on a thin conical brane and
their extension into a five and six-dimensional bulk with a
Gauss-Bonnet term were found~\cite{CPTZ}.

Moreover we are still lacking an understanding of time dependent
cosmological solutions in codimension-2 braneworlds. To have a
cosmological evolution we need regularized branes, the brane
world-volume should be expanding and in general the bulk space
should also evolve in time. This is a formidable task, so we
follow an alternative approach
\cite{Papantonopoulos:2007fk,Minamitsuji:2007fx}.
\footnote{Recently, some works have studied the cosmological
evolution of codimension-1 or 2 branes in six dimensional models
\cite{codim1&2}}

First for our setup we consider another way of regularization
which was recently proposed and consists merely of the reduction
of codimensionality of the brane. In this approach, the bulk
around the codimension-2 brane is cut close to the conical tip and
it is replaced by a codimension-1 brane which is capped by
appropriate bulk sections \cite{PST} (see \cite{Gott} for a
similar regularization of cosmic strings in flat spacetime). This
regularization has been applied to flux compactification systems
in six dimensions for unwarped ``rugby-ball''-like solutions in
\cite{PST}. The case of warped solutions with conical branes (with
or without supersymmetry) \cite{ppz} and even more general warped
solutions allowing non-conical branes \cite{tas} have been
studied. More precisely we consider the case of one capped
regularized conical brane of codimension one and give the {\it
static} bulk and cap solutions.

Then we study the motion of the regularized codimension-1 brane in
the space between the bulk and the brane-cap which remains static
(see \eg \cite{Cuadros-Melgar:2005ex}).  In this way, a
cosmological evolution will be induced on the brane in a similar
way as in the mirage cosmology \cite{mirage}, but with the
inclusion of the back-reaction of the brane energy density (\ie
the brane is not considered merely a probe one). Since in the
mirage cosmology, the four-dimensional scale factor descends from
the warp factor in the four-dimensional part of the bulk metric,
we will discuss the regularized brane in the case of warped bulk
\cite{ppz}, rather than unwarped bulk \cite{PST}. It is worth
noting that the above procedure provided in five dimensions the
most general isotropic brane cosmological solutions
\cite{5dcosmo}.

We find the Friedmann equation on the brane by solving the Israel
junction conditions, which play the r\^ole of the equations of
motion of the codimension-1 brane. We find that at early times
cosmology is dominated by an energy density term proportional to
$\rho^2$, like in the Rundall-Sundrum model in five dimensions.
However, at late times where the brane moves close to its
equilibrium point, which in turn is close to the would-be conical
singularity, the coefficient of the linear to the energy density
term is negative (\ie we obtain negative effective Newton's
constant). Thus, we cannot recover the standard four dimensional
cosmology at late times. This seems to be the consequence of
considering the bulk sections static. It is possible, that this
behaviour is due to a ghost mode appearing among the perturbations
of the system, after imposing the staticity of the bulk sections.
Furthermore, the above result points out that there is a
difference  between the six-dimensional brane cosmology in
comparison to  the five-dimensional one. The study of brane
cosmology in Einstein gravity in five dimensions, can be made
either in a gauge where the bulk is time-dependent and the brane
lies at a fixed position, or in a gauge where the bulk is static
and the brane  movement into the bulk induces a cosmological
evolution on it \cite{equivalence}. This, however, does not seem
to hold in six-dimensions anymore.

The talk is organized as follows. In Sec.~2 we present the static
regularized brane solution in a bulk of general warping. In Sec.~3
we derive the equations of motion of the codimension-1 brane and
in Sec.~4 we study the induced cosmological evolution on the
moving brane. Finally in Sec.~5 we draw our conclusions.

\section{Regularized static brane solutions}

\begin{figure}[t]
\begin{center}
\begin{picture}(200,160)(-60,-20)

\Text(45,155)[c]{$r=+1$} \Text(45,-10)[c]{$r=-1$}

\Text(-65,72)[c]{Bulk: $c_0,~R_0,~\a$}

\Text(125,130)[c]{Cap: $c_c,~R_c,~\a$}

\Text(-65,142)[c]{4-brane} \Text(-65,129)[c]{$\la,~v$}
\Text(-65,114)[c]{$r=r_c$}

\BBoxc(-65,-5.5)(50,30)

\Text(-65,2)[c]{3-brane} \Text(-65,-13)[c]{$T$}

\Vertex(42,2){2}

\Oval(-65,128)(22,38)(0)

\LongArrow(-20,125)(25,125) \LongArrow(-30,2)(33,2)

\epsfig{file=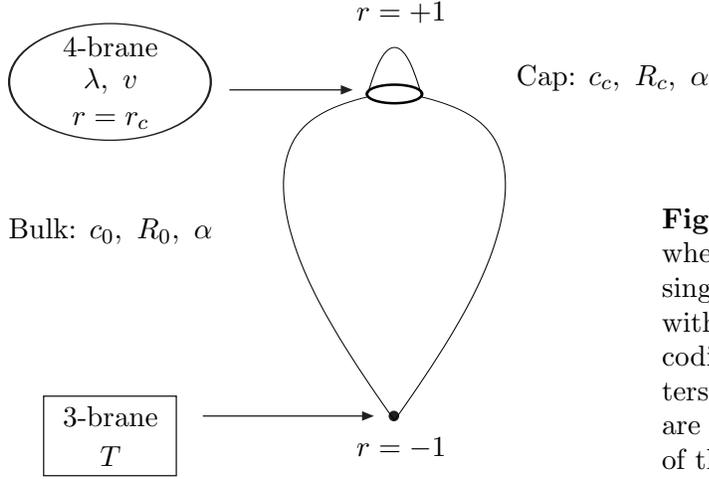,width=3cm,height=5cm}

\end{picture}
\begin{minipage}[b]{14pc}
\caption{ The internal space where the upper codimension-2
singularity has been  regularized with the introduction of a
ring-like codimension-1 brane. The parameters of the action and
the solution are denoted in the appropriate part of the internal
space.} \label{internalfig}
\end{minipage}
\end{center}
\end{figure}

In this section we will briefly give the necessary results of the
static solution which we will need in the following for the brane
motion. The bulk theory that we use is a six-dimensional
Einstein-Maxwell system which in the presence of a positive
cosmological constant and a gauge flux, spontaneously compactifies
the internal space \cite{spontan}. The known axisymmetric
solutions have in general two codimension-2 singularities at the
poles of a deformed sphere \cite{japs}. We study the case where
only one (\eg the upper) codimension-2 brane is regularized by the
introduction of a ring-like brane at $r=r_c$ with an appropriate
cap. The dynamics of the system is given by the following action
\be S= \int d^6 x \sqrt{-g} \left( {M^4 \over 2} R - \La_i -{1
\over 4}{\cal F}^2 \right) - \int d^5 x \sqrt{-\g_+}\left(\la
+{v^2 \over 2} (\tilde{D}_{\hat{\m}} \s)^2 \right)-\int d^4x
\sqrt{-\g_-} ~T~, \ee where $M$ is the six-dimensional fundamental
Planck mass, $\La_i$ are the bulk ($i=0$) and cap ($i=c$)
cosmological constants, ${\cal F}_{MN}$ the gauge field strength,
$T$ the tension of the lower codimension-2 brane, $\la$ the
4-brane tension, $\s$ the 4-brane Goldstone scalar field necessary
for the regularization and $v$ the vev of the Higgs field from
which the Goldstone field originates. For the coupling between the
Goldstone field and the bulk gauge field we use the notation
$\tilde{D}_{\hat{\m}} \s = \de_{\hat{\m}} \s -e~ a_{\hat{\m}}$,
with $a_{\hat{\m}}={\cal A}_M \de_{\hat{\m}} X^M$ the pullback of
the  gauge field on the ring-like brane and $e$ its coupling to
the scalar field. In  the above action we omitted the
Gibbons-Hawking term. The configuration is shown in more detail in
Fig.\ref{internalfig}.

The solution for the bulk and cap regions depends on a parameter
$\a$ which is a measure of the warping of the space (for $\a=1$ we
obtain the unwarped case) and is given by \cite{ppz} \bea
ds_6^2&=&  z^2 \eta_{\m\n} dx^\m dx^\n + R_i^2\left[ {dr^2 \over f} +c_i^2 f~ d\vf^2 \right]~,\\
{\cal F}_{r \vf}&=&  -c_i R_i M^2 S \cdot {1 \over z^4}~, \eea
with $R_i^2=M^4/(2 \La_i)$ and the following bulk functions \bea
z(r)&=&{1 \over 2}[(1-\a)r+(1+\a)]\\ f(r)&=& {1 \over 5
(1-\a)^2}\left[ -z^2 + {1-\a^8 \over 1-\a^3}\cdot{1 \over z^3
}-\a^3{1-\a^5 \over 1-\a^3}\cdot{1 \over z^6} \right]~, \eea with
$S(\a)= \sqrt{{3 \over 5}\a^3{1-\a^5 \over 1-\a^3}}$. The range of
the internal space coordinates is $-1 \leq r \leq 1$ and $0\leq
\vf <2 \pi$. Taking into account that in the limit $r \to \pm 1$,
it is $f \to 2(1\mp r)X_\pm$ with the constants $X_\pm$ given by
\be X_+= {5+3\a^8-8\a^3 \over 20 (1-\a)(1-\a^3)} ~~~ ,~~~ X_-=
{3+5\a^8-8\a^5 \over 20 \a^4(1-\a)(1-\a^3)} ~, \ee the cap is
smooth at $r=+1$ as long as is $c_c=1/X_+$. Furthermore, the
metric is continuous if $c_0 R_0=c_c R_c$, which gives $R_c= \b_+
R_0$ with $\b_+=X_+ c_0$ \footnote{In this brief presentation of
the background, we have taken $\xi=1$ in comparison with
\cite{ppz}. The physical quantities, however, are
$\xi$-independent and depend only on $\b_+$.}. The conical
singularity at $r=-1$ is supported by a codimension-2 brane with
tension \be T=2\pi M^4 (1- c_0 X_- ) \ee while the parameters of
the $4$-brane $\la$, $v$ are fixed by the radii $R_0$, $R_c$ and
the brane position $r_c$ \cite{ppz}. Finally, the gauge field is
quantized as \be 2 c_0 R_0 M^2 e~Y = N   ~~~,~~~N \in \mathbb{Z}~,
\label{N} \ee with $Y={(1-\a^3) \over 3 \a^3 (1-\a)}~S$ and the
brane scalar field has  solution $\s= n \vf$ with $n \in
\mathbb{Z}$. The two quantum numbers $n$, $N$ are related through
the junction conditions as
 \be n={N
\over 2}{2 \over (1-\a^3)}\left[{5(1-\a^8) \over 8 (1-\a^5)}
-\a^3\right]~. \ee
 Since the quantities $n$, $N$ are integers,
the above relation imposes
 a restriction to the values of the admissible warpings $\a$, which implies that static solutions are
consistent only for discrete values of the warping $\a$.

\section{The {\bf cosmological }dynamics of the $4$-brane}

To study the cosmological evolution on the brane we include matter
on it which in general makes both the brane and the bulk to evolve
in time. Unfortunately we do not know any time dependent solution
for such setup. Therefore we will approach the problem in a
simplest way, by making the approximation that the bulk remains
{\it static} and that the brane matter merely makes the brane to
move between the static bulk and the static cup, away from its
equilibrium point $r = r_c$, thus obtaining cosmology on the
brane, as in the mirage cosmology approach \cite{mirage}. This is
to be regarded as a first step towards understanding the generic
brane cosmological evolution. In our case the brane is not merely
a probe brane and we will use the junction conditions to derive
the induced cosmology, thus taking into account the back-reaction
of the brane energy density.

Firstly, to embed  the brane in the static bulk, let us take  the
brane coordinates  be $\s^{\hat{\m}}=(\s,x^i,\vf)$.  [The
brane-time coordinate $\s$ is not to be confused with the
Goldstone field $\s$ which will not appear in our subsequent
analysis.] Then the brane embedding $X^M$ in the bulk is taken for
both sections to be \be X^i=x^i ~~~ ,~~~ X^r={\cal R}(\s)~~~ {\rm
and}  ~~~ X^\vf=\vf~, \ee while for the time coordinate embedding
we choose for the outer  bulk  section \be X^0_{(out)}=\s~, \ee
and for the inner cap section \be X^0_{(in)}=T(\s)~. \ee

The continuity of the induced metric $\g_{\hat{\m} \hat{\n}} =
g_{MN} \de_{\hat{\m}} X^M \de_{\hat{\n}} X^N $, apart from the
relation $c_0 R_0= c_c R_c$ as in the static case, gives a
relation of the time coordinate $T$ in the upper cap region with
the brane time coordinate $\s$ (dots are with respect to $\s$) \be
\dot{T}^2 \left(1 - \b_+^2{\dot{\cal R}^2 \over \dot{T}^2} {R_0^2
\over f z^2} \right)= \left(1 - \dot{\cal R}^2 {R_0^2 \over f z^2}
\right) ~.\label{T} \ee Then the induced metric $\g_{\hat{\m}
\hat{\n}}$ on the brane reads \be ds^2_{(5)}=
-z^2\left(1-{\dot{{\cal R}}^2{R_0^2 \over f
z^2}}\right)d\s^2+z^2d\vec{x}^2+c_0^2R_0^2 f d\vf^2 ~. \ee

The continuity of the gauge field, on the other hand, is
guaranteed by the fact that its only non-vanishing component is
$A_\vf$ and $X^\vf$ is $\s$-independent.

Secondly, we introduce an energy momentum tensor of a perfect
fluid on the brane $t_{\hat{\m}}^{\hat{\n}
~(br)}=-(2/\sqrt{-\g_+}) \d S_{br}/ \d \g^{~ \m}_{+ ~ \n} = {\rm
diag}(-\r, P,P,P,\hat{P})$ (where $S_{br}$ is the brane action).
We also consider a possible coupling of the brane matter to the
bulk gauge field (consistent with the cosmological symmetries) by
$\d S_{br} / \d a^{\hat{\k}}=(l,L,L,L,\hat{L})$. Splitting the
above quantities to one part responsible for the static solution
and another expressing the presence of matter on the brane, we
have
 \bea
\r&=&\la + {v^2 (n-e {\cal A}_\vf^+)^2 \over 2 c_0^2 R_{0}^2  f(r_c)} + \r_m \equiv  \r_0+ \r_m \label{rsplit} \\
P&=&-\la - {v^2 (n-e {\cal A}_\vf^+)^2 \over 2 c_0^2 R_{0}^2  f(r_c)} +P_m \equiv  -\r_0+ P_m\\
\hat{P}&=&-\la + {v^2 (n-e {\cal A}_\vf^+)^2 \over 2 c_0^2 R_{0}^2  f(r_c)} + \hat{P}_m\\
l&=&l_m\\
L&=&L_m \\
\hat{L}&=& e  v^2 (n-e {\cal A}_\vf^+)+\hat{L}_m \eea with all the
other quantities vanishing and $\r_m$, $P_m$, $\hat{P}_m$, $l_m$,
$L_m$, $\hat{L}_m$ the matter contributions.

In the spirit of mirage cosmology approach \cite{mirage} the brane
matter merely makes the brane to move in the static bulk. As we
have already stated this movement induces cosmology on the brane
and warping of the bulk is necessary. Therefore it is convenient
to rewrite the metric in the form \be ds^2_{(5)}= -d\t^2+a^2(\t)
d\vec{x}^2+ b^2(\t) d\vf^2 ~, \ee with $a= z({\cal R}(\t))$ and
$b= c_0 R_0 \sqrt{f({\cal R}(\t))}$. The brane proper time is
given by \be \dot{\t}^2=z^2\left(1-{\dot{{\cal R}}^2{R_0^2 \over f
z^2}}\right)~. \label{proper} \ee From now on we will assume
without loss of generality that $\dot{\t}>0$.  It is evident from
the above, that cosmological evolution from mere motion of the
brane in the static bulk is possible only when there is warping in
the bulk. In the unwarped version of this model \cite{PST}, mirage
cosmology is impossible and some bulk time-dependence is
compulsory.

The Hubble parameters for the two scale factors are given by
 \be
H_a \equiv {1 \over a}{ d a \over d \t } =  {z' \over z^2} {
\dot{{\cal R}} \over \sqrt{ 1-{\dot{{\cal R}}^2{R_0^2 \over f
z^2}}  } }~, \label{Ha} \ee
 and
 \be H_b \equiv {1 \over b}{ d b \over d \t } =  {f' \over 2 f z} {
\dot{{\cal R}} \over \sqrt{ 1-{\dot{{\cal R}}^2{R_0^2 \over f
z^2}}  } }~. \ee Then the  ratio of $H_a$ and $H_b$ gives a
precise relation between them for our particular model. It is
given by
 \be
H_b={z f' \over 2 f z'} H_a~, \ee and we notice that since in the
model we study it is always  $f' <0$, in the neighborhood of
$r=1$, the two Hubble parameters have opposite sign. This means
 that if the four dimensional space expands, the internal space shrinks.

Apart from the continuity conditions we have to take into account
the junction conditions for the derivatives of the metric and the
gauge field, which read \bea
\{\hat{K}_{\hat{\m}\hat{\n}}\}&=&-{1 \over M^4} t_{\hat{\m} \hat{\n}}^{(br)}~, \label{Kjunction} \\
\{n_{M} {\cal F}^{M}_{~N}\de_{\hat{\k}} X^N  \}&=& - {\d S_{br}
\over \d a^{\hat{\k}}}   ~. \label{Fjunction} \eea We denote
$\{H\}=H^{in} + H^{out}$ the sum of the quantity $H$ from each
side of each brane. The extrinsic curvatures are constructed using
the normal to the brane $n_M$ which points {\it inwards to the
corresponding part of the bulk} each time (we use the conventions
of \cite {CR}). \footnote{The left hand sides of the above
equations are computed in detail in the Appendix A of
\cite{Papantonopoulos:2007fk}}.

So far we have $6$ parameters: $\r, P, \hat{P}, l, L, \hat{L}$ and
with the use of the continuity relations of the induced metric,
$c_0 R_0= c_c R_c$ and (\ref{T}), the $(t)$ and $(i)$ components
of the junction conditions for the gauge field give for the
coupling: $l = L = 0$, the $(\vf)$ component of the gauge field
junction and the $(\vf\vf)$ component of the metric junction
determine $\hat{P}$ and $\hat{L}$. Finally the two remaining
metric junction conditions ($(\s\s)$ and $(ij)$ components) give
the Friedmann equation and the acceleration equation respectively.

The obtained Friedmann equation is \be H_a^2 = {1 \over 4M^8 {\cal
C}^2 {\cal A}}~\r^2 + {M^8 {\cal C}^2 (1-\b_+^2)^2 \over 4 R_0^4
\b_+^4 {\cal A}}\cdot{1 \over \r^2}-{1+\b_+^2 \over 2 \b_+^2 R_0^2
{\cal  A}}~ \label{fried} \ee where ${\cal A}={ z^2 \over
fz^{'2}}$ and ${\cal C}=\sqrt{f}\left( 3 {z' \over z}+{f' \over
2f}\right)$, both evaluated on the brane. It has unconventional
dependence on the energy density, in particular the inverse square
dependence is known to occur for motions in backgrounds of
asymmetrical warping \cite{asym}.

The equilibrium point $r_c$ of the system is found if we set
$H_a=0$, which gives the brane energy density without matter \be
\r_0=-{M^4 {\cal C}_c \over R_0 \b_+}(1- \b_+)~, \ee where ${\cal
C}_c$ is the value of ${\cal C}$ at $r_c$  and is the same
appearing  in \reef{rsplit}. The behaviour of the function ${\cal
C}$ is that, as $r \to -1$ it limits to ${\cal  C} \to  \infty$
and monotonically decreases and limits to ${\cal  C} \to  -\infty$
as $r \to 1$. On the other hand ${\cal A}$, is always positive
with ${\cal A} \to \infty$ as $r \to \pm 1$, and ${\cal O}(1)$ in
the intermediate region. In the unwarped limit $\a \to 1$ the
Friedmann equation becomes as expected trivial, \ie $H_a=0$.

Before studying various limits of the above equation, let us
define the effective four dimensional matter energy density
$\r_m^{(4)}$ by averaging over the azimuthal direction (we assume
that $\r_m$ is independent of $\vf$) \be \r_m^{(4)}=\int d\vf
\sqrt{g_{\vf \vf}} \r_m = {2 \pi \b_+ \over X_+} R_0 \sqrt{f} \r_m
~, \ee with similar definitions for the other 4-brane quantities.

Let us suppose now that initially $-1 \ll {\cal R}(\s) < r_c < 1$,
with $1-r_c \ll 1$. The goal is to find how ${\cal R}(\s)$
behaves.  To recover a four-dimensional Friedmann equation at late
times we can assume that the brane energy density is small in
comparison with the static case energy density, \ie $\r_m^{(4)}
\ll \r_0$,  so we can expand \reef{fried} in powers of
$\r_m^{(4)}$ and obtain the following four dimensional form of the
Friedmann equation \be H_a^2 = {8 \pi \over 3} G_{eff} \r_m^{(4)}
+ \D(a) + {\cal O}(\r_m^{(4)~2})~, \ee where the effective
Newton's constant is \be G_{eff}={3X_+ {\cal C}_c (1-\b_+) \over
32 \pi^2 R_0^2 \b_+^2 M^4 {\cal A}~{\cal C}^2 \sqrt{f}} \left[
{{\cal C}^4 \over {\cal C}_c^4}\left({1 + \b_+ \over 1 -
\b_+}\right)^2 -1\right]~. \ee The quantity $\D(a)$ depends on the
parameters of the bulk and  plays the r\^ole   of the the mirage
matter induced on the brane from the bulk and it is given by \be
\D(a)={ (1- \b_+)^2 \over 4 R_0^2 \b_+^2 {\cal A} } \left[ { {\cal
C}_c^2 \over {\cal C}^2 }+{ {\cal C}^2 \over {\cal C}_c^2
}\left({1 + \b_+ \over 1 - \b_+}\right)^2 -2 { 1+ \b_+^2 \over
(1-\b_+)^2} \right] ~. \ee

\begin{figure}[t]
\begin{center}
\begin{picture}(200,130)(0,0)

\LongArrow(-20,50)(210,50) \LongArrow(80,0)(80,130)

\Vertex(170,50){1.5} \Vertex(-10,50){1.5}

\Vertex(107.5,50){2} \Vertex(150,50){2}

\DashLine(170,70)(170,0){2} \DashLine(-10,70)(-10,0){2}
\DashLine(107.5,50)(107.5,120){2}

\Curve{(110,120)(130,53)(135,50)(140,47)(165,0)}
\Curve{(-5,0)(30,35)(80,46)(85,50)(105,120)}

\Text(180,60)[c]{$+1$} \Text(5,60)[c]{$-1$}

\Text(205,60)[c]{${\cal R}(\s)$} \Text(150,60)[c]{$r_c$}
\Text(95,140)[c]{$G_{eff}$} \Text(107.5,38)[c]{$r_d$}

\end{picture}\hspace{1.8pc}
\begin{minipage}[b]{14pc}
\caption{ The generic form of the effective Newton's constant
$G_{eff}$  as a function of the brane position ${\cal R}(\s)$. As
the brane approaches the equilibrium point $r_c$, we always have
$G_{eff}<0$. Additionally, $G_{eff}$ diverges as $r \to \pm 1$ and
at one point $r_d$ in between.} \label{Geffpic}
\end{minipage}
\end{center}
\end{figure}
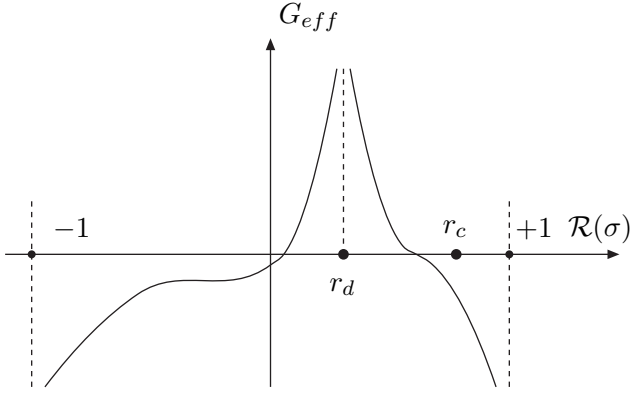

\begin{figure}[t]
\begin{center}
\begin{picture}(200,130)(0,0)

\LongArrow(-20,20)(210,20) \LongArrow(80,0)(80,130)

\Vertex(170,20){1.5} \Vertex(-10,20){1.5}

\Vertex(107.5,20){2} \Vertex(150,20){2}

\DashLine(170,100)(170,20){2} \DashLine(-10,100)(-10,20){2}
\DashLine(107.5,20)(107.5,120){2}

\Curve{(110,120)(140,10)(150,20)(170,30)}
\Curve{(-10,50)(35,20)(70,20)(105,120)}

\Text(180,30)[c]{$+1$} \Text(-20,30)[c]{$-1$}

\Text(205,30)[c]{${\cal R}(\s)$} \Text(157,8)[c]{$r_c$}
\Text(95,140)[c]{$\D$} \Text(107.5,8)[c]{$r_d$}

\end{picture} \hspace{1.8pc}
\begin{minipage}[b]{14pc}
\caption{ The generic form of the mirage matter contribution $\D$
as  a function of the brane position ${\cal R}(\s)$. At the static
equilibrium point $r_c$ it vanishes and at $r_d$ it diverges.}
\label{Dpic}
\end{minipage}
\end{center}
\end{figure}
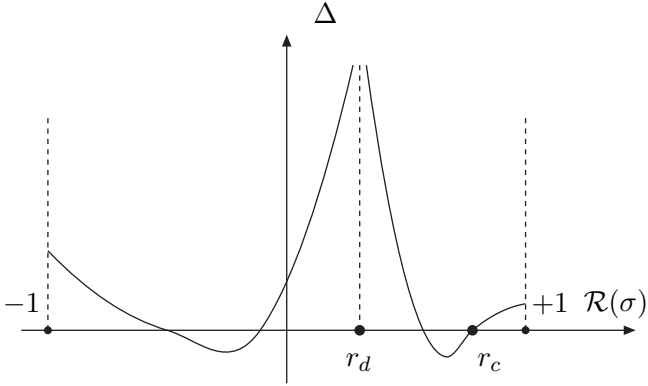

The behaviour of $G_{eff}$ as a function of ${\cal R}(\s)$ is
given  in Fig.~\ref{Geffpic} and has the following important
features: At the points where the geometry becomes conical ($r \to
\pm 1$) the effective Newton's constant is negative and diverging.
In between, there is a point $r_d$, with
 \be z_d = \left({3 (1-\a^8) \over 8 (1-\a^3)} \right)^{1/5}
~, \ee which is a root of ${\cal C}$ and $G_{eff}$ diverges to $+
\infty$. At this point the matter energy density is bound to
vanish. It is important to note that even in the region where
$G_{eff}$ is positive, there is always a strong time variation of
$G_{eff}$, which for \be {1\over G_{eff}}{d G_{eff} \over d\t}=
H_a \d ~, \ee has $\d > {\cal O}(10)$, in contradiction with
observations \cite{obs} which dictate that $\d < 0.1$. But even
more important is the fact that close to the static equilibrium
point, where the cosmology is supposed to mimic best the one of a
codimension-2 brane, we get {\it negative} Newton's constant \be
G_{eff}({\cal R}=r_c)= {3X_+ \over 8 \pi^2 R_0^2 \b_+ (1-\b_+) M^4
{\cal A}~{\cal C}_c \sqrt{f}}   <0~, \ee since we have that for
any value of the parameters and for $r_c$ in the neighborhood of
$r=+1$, it is ${\cal C}_c <0$.

Let us also look at the mirage matter contribution $\D$, which  is
depicted in Fig.~\ref{Dpic}. As expected, it vanishes for the
static  equilibrium point $r_c$, it is finite at the boundaries
$r=\pm 1$ and diverges at the root $r_d$ of ${\cal C}$. Again
there is no region in the brane position interval where the
contribution of $\D$ is constant enough to resemble a cosmological
constant contribution to the four dimensional Friedmann equation.

This result is not altered by supersymmetrizing the model
\cite{susy}. The bulk and cap solutions are different from the
non-supersymmetric case due to the presence of the dilaton.
However, the only difference in the Friedmann is a redefinition of
the quantities ${\cal A}$ and ${\cal C}$. In the supersymmetric
case, these quantities read \be {\cal A}_{susy}=4{ z^2 \over
fz^{'2}} ~~~{\rm and}~~~{\cal C}_{susy}=\sqrt{f}\left( {3z' \over
2z}+{f' \over 2f}\right)~.\ee It is easy to see again that the
effective Newton's constant is negative for the motion near the
would-be conical tip and that even away from that point, it is
very strongly varying.

On the opposite limit, that the matter energy density is much
larger  than the static case energy density, \ie $\r_m^{(4)} \gg
\r_0$, we get the expected asymptotics \be H_a^2 = {1 \over 4M^8
{\cal C}^2 {\cal A}}~\r_m^2~, \ee which is a five-dimensional
Friedmann law (with time-varying five-dimensional Newton's
constant) at early times.

Taking under consideration the difficulties of the model to give a
Fiedmann equation with the correct sign of $G_{eff}$, we will not
proceed with the presentation of the analysis of the acceleration
equation. With this equation, one finds even more difficulties
towards obtaining a realistic four-dimensional evolution. For
example, in the low energy limit, one gets a coefficient of the
linear energy density term, which is not related to the $G_{eff}$,
that we obtained from the Friedmann equation, in the way it does
in standard four-dimensional General Relativity.

This pathological features of the low energy density limit, where
the  expansion does not appear to have an effective
four-dimensional limit, can find a potential explanation when
looking at the energy continuity equation on the brane. Taking the
covariant divergence of the Israel junction  condition
(\ref{Kjunction}), together with the Codazzi equation
$\nabla^{(4)}_{\hat{\m}} \hat{K}^{\hat{\m}}_{\hat{\n}}=G_{K
\La}n^{\La} h_N^K \de_{\hat{\n}} X^N ~$ and the bulk Einstein
equation, we arrive at the following simple equation \cite{CR} \be
\nabla^{(4)}_{\hat{\m}} t^{\hat{\m}~(br)}_{\hat{\n}} = - \{
T^{(B)}_{K \La}h^K_N n^{\La} \de_{\hat{\n}}X^N  \small{\}}~. \ee

Because of the jump of the bulk energy momentum tensor across the
4-brane, the energy-momentum tensor on the 4-brane is not
conserved. This is a usual feature of moving brane cosmologies in
asymmetrically warped backgrounds \cite{asy}. In more detail, the
form of the above equation for the particular model is given by
 \be
{d \r \over d\t}+ 3(\r + P) H_a +(\r + \hat{P}){z f' \over 2 f z'}
H_a = - {S^2 \r H_a \over z' z^7 {\cal C} \sqrt{f}}~, \ee where in
the right hand side we have used the Friedmann equation
\reef{fried}. In a straightforward but lengthy calculation, one
can see that the latter  equation can be derived from the $(\s\s),
(ij)$ and $(\vf\vf)$ components of the junction conditions of
(\ref{Kjunction}). The problems in the four-dimensional limit,
that we faced previously, can be traced to large energy
dissipation off the brane as well as a large work done during the
contraction of the ring-brane.

\section{Conclusions}

In the present presentation we made a first step towards the study
of the cosmology of a codimension-2 brane. We regularized the
codimension-2 singularities by the method of lowering its
codimensionality. We cut the space close to the conical tip and
replaced it by a ring brane with an appropriate cap. Unfortunately
we do not know any time dependent solution for such setup.
Therefore we will approach the problem in a simplest way, we
assumed that the bulk and the cap remain static as the brane moves
between them. The motion of the brane then induces a cosmological
evolution for the matter on the brane. The junction conditions
provide the Friedmann and acceleration equations on the brane.

From the Friedmann equation we can see that we cannot recover
standard cosmological evolution of the brane at low energies. The
effective Newton's constant is negative in the interesting limit
that the brane approaches its equilibrium point, close to the
would-be conical singularity. In other words, we obtain
antigravity in this limit. Even away from this point, \ie when the
brane moves away from its equilibrium point, the Newton's constant
varies significantly in contradiction with standard cosmology. At
one position of the internal space, the Newton's constant even
diverges and forces the matter energy density to vanish. Taking
all the above  into account, we did not present the further
analysis of the system by considering the acceleration equation.

The reason for this unconventional cosmological evolution, is the
use of the specific restricted ansatz for the solution of the
system's equation of motion. The staticity of the bulk was proved
to be an oversimplification. We imagine that this restriction on
the system may result to the appearance of some scalar mode in the
perturbative analysis of  \cite{stabwarp}, which for a certain
region of the brane motion (close to the pole of the internal
manifold) is ghost-like. This mode may then be responsible for the
negative effective Newton's constant. The unrestricted
perturbative analysis in \cite{stabwarp} resulted in a linearized
four-dimensional Einstein equation for distances larger than the
compactification scale. The same happened in \cite{PST} in the
unwarped model perturbation analysis\footnote{The question of
stability of the regularized models was not discussed in
\cite{PST,stabwarp} and it could be that these compactifications
have modes with negative mass squared.} (see \cite{kaloper} for a
related analysis with a brane induced gravity term), even though
it would not have mirage evolution, as we studied it here, because
of the absence of a warp factor. The gradient expansion technique
was recently used in similar six dimensional models
\cite{gradient} to derive a low energy effective theory on the
regularized brane and show that standard four-dimensional Einstein
gravity is recovered at low energies. Consequently, this approach
could be used in our case to derive the low energy effective
theory \cite{PPZgradient}. The appearance of the standard
four-dimensional linearized dynamics, shows that they are realized
when the bulk is necessarily time-dependent.

Clearly, the next step should be the study of the system in a
setup where the bulk is also time-dependent. In that respect,
brane cosmology in six dimensions seems to be different for the
one in five dimensions. In Einstein gravity in five dimensions,
one can always work in a gauge where the bulk is static and the
brane acquires a cosmological evolution by moving into the bulk.
However, in six dimensions there are more degrees of freedom which
make this gauge choice not general. In the present paper, we have
frozen these degrees of freedom hoping to find  consistent
cosmological solutions, but it turned out that this does not give
a viable cosmology. There are several known time-dependent
backgrounds which can be used to look for realistic brane
cosmologies \cite{timedep}. We plan to address this issue in the
near future.

\ack

This talk is based on the works done in collaboration with E.
Papantonopoulos and A. Papazoglou. We thank the organizers of the
13th NEB Conference, held in Thessaloniki, for their kind
hospitality and perfect organization. This work was supported by
the Hellenic Army Academy and the NTUA research program PEVE07.

\end{document}